\documentclass[twocolumn,aps,floatfix,showpacs,showkeys,prl,superscriptaddress,preprintnumbers]{revtex4-1}
\usepackage{epsfig}
\usepackage{graphicx}
\usepackage{amsmath}
\usepackage{tabularx}

\begin{document}
\title{Orbital magnetic moment and extrinsic spin Hall effect for iron
  impurity in gold}
\author{Alexander B. Shick}
\affiliation{Institute of Physics, Academy of Sciences of the Czech
Republic, Na Slovance 2, CZ-182 21 Prague, Czech Republic}

\author{Jind\v{r}ich Koloren\v{c}}
\affiliation{University of Hamburg, Jungiusstra\ss e 9, 20355 Hamburg,
Germany}
\affiliation{Institute of Physics, Academy of Sciences of
the Czech Republic, Na Slovance 2, CZ-182 21 Prague, Czech Republic}

\author{V\'{a}clav Jani\v{s}}
\affiliation{Institute of Physics, Academy of Sciences of
the Czech Republic, Na Slovance 2, CZ-182 21 Prague, Czech Republic}

\author{Alexander I. Lichtenstein}
\affiliation{University of Hamburg, Jungiusstra\ss e 9, 20355 Hamburg,
Germany}

\date{\today}

\begin{abstract}
We report electronic structure calculations of an iron impurity in
gold host.  The spin, orbital and dipole magnetic moments were
investigated using the LDA+$U$ correlated band theory. We show that
the {\em around-mean-field}-LDA+$U$ reproduces the XMCD experimental
data well and does not lead to formation of a large orbital moment on
the Fe atom. Furthermore, exact diagonalization of the multi-orbital
Anderson impurity model with the full Coulomb interaction matrix and
the spin-orbit coupling is performed in order to estimate the spin
Hall angle. The obtained value $\gamma_S \approx 0.025$ suggests that
there is no giant extrinsic spin Hall effect due to scattering on iron
impurities in gold.
\end{abstract}


\maketitle


During the last several years, a broad interest and attention have
been devoted to the spin Hall effect (SHE) in
semiconductors~\cite{Jungwirth2010} and metals~\cite{Kimura2007}.
This effect amounts to an observation of a transversal spin current
when a charge current is flowing through a solid. The SHE is caused by
the spin-orbit coupling (SOC) and can occur even in non-magnetic
solids~\cite{Kimura2007}.

Recently, an experimental observation of a giant  SHE in Au/FePt has been reported~\cite{Seki}.
The spin Hall conductivity of $\sim$ $10^5$ $\Omega^{-1}$cm$^{-1}$ and the spin Hall angle   as
large as $\sim$ 0.1 were measured~\cite{Seki}. 
Guo {\em et al.} suggested the effect to be of extrinsic origin due to the Fe and Pt impurities 
in  gold ~\cite{Guo}. They reported local (spin) density approximation (LSDA)
plus Coulomb $U$ (LDA+$U$)  solution for Fe in Au with a very large orbital magnetic moment $M_L\sim 1.5 \mu_B$. 

The results of Ref.~\cite{Guo} contradict the LDA+$U$ calculations presented in
Ref.~\cite{Costi} that reported a tiny
$M_L\sim 0.02\mu_B$.  The value of $M_L$ from  Ref.~\cite{Guo}  is clearly inconsistent with the 
experimental x-ray magnetic circular dichroism (XMCD)
data for the ratio
of $M_L$ and the effective spin moment $M_S$, $R_{LS} = 0.034$~\cite{Brewer}. Assuming  $M_S \sim 3 \mu_B$ 
leads to  $M_L \sim 0.1 \mu_B$ which is of an order  of magnitude smaller
than the prediction of Ref.~\cite{Guo}. It was suggested \cite{BoGu}
that the discrepancy between  Refs.~\cite{Guo} and
\cite{Costi} 
are due to  different choices of the Coulomb $U$. A large value of $M_L$ is calculated with 
$U = 5$ eV \cite{Guo},
while much smaller ones are obtained with $U = 3$ eV~\cite{Costi}.

In this work we revisit the electronic and magnetic structure of the Fe impurity in Au. 
We examine different flavors of the rotationally-invariant LDA+$U$ method~\cite{Lixt95}: 
the ``fully localized limit'' (FLL) as well as the ``around mean field'' (AMF) version. 
The results for the orbital magnetic moment $M_L$ are compared with
the available experimental data.  

Both LSDA and LDA+$U$ methods yield broken-symmetry static mean-field 
solutions with ordered spin and orbital moments, whereas the true dynamical solution of an
impurity in a non-magnetic host exhibits $M_S=2\langle\hat S_z\rangle=0$ and $M_L=\langle\hat L_z\rangle=0$ when no external magnetic 
field is applied and no preferential direction for the
orientation of the moments exists. In order to go beyond the static mean-field  and to incorporate the  dynamical electron
correlations, we employ the exact diagonalization (ED) method to solve
a multi-orbital single impurity Anderson model (SIAM)~\cite{Hewson} whose parameters are 
extracted from LDA calculations.
We evaluate the spectral density at the Fe impurity in Au and estimate
the spin Hall angle due to skew scattering on the impurity. A
relation between the electronic structure and the extrinsic SHE is discussed.

As a computational model we use a FeAu$_{15}$ supercell 
chosen to keep Fe and its 12 nearest Au neighbors separated 
from other impurity atoms.  No relaxation is performed as  it is not essential for the closed
packed {\em fcc} structure. We use the lattice constant of elemental Au,
$a=7.71$~a.u. All calculations are performed making use of a relativistic version (with SOC) 
of LDA+$U$ implemented in the linearized augmented plane wave  (FP-LAPW) basis~\cite{shick01}. 
The radii of the atomic muffin-tin
(MT) spheres are set to 2.3~a.u.~(Fe) and 2.5~a.u.~(Au). The parameter $R
\times K_{\rm max}=7.6$ determined the basis set size and the
Brillouin zone was sampled with 343 $k$~points. We
checked that a finer sampling with 729 $k$~points
does not modify the results. 

First, we apply the conventional LSDA with the von Barth and Hedin~\cite{vBH}  exchange-correlation potential
 implemented within the relativistic FP-LAPW  method~\cite{shick97}.
Our results  for
the spin and orbital moments inside the Fe MT sphere, $M_S$ and $M_L$, are
compared with the results of other calculations in
Table~\ref{tab:1}. In spite of  a relatively small (16 atoms only)
and unrelaxed supercell, present results are in a fair agreement with
VASP~\cite{Costi} results for a substantially
larger and relaxed supercell containing 108 atoms, with the tight-binding LMTO (TB-LMTO) results for 55 atom
supercell~\cite{Frota}, and with the KKR-ASA
calculations~\cite{Chadov}. All calculations indicate a small value of $M_L$
for Fe impurity in  Au, 
which is typical for $3d$ transitional metals and alloys. The calculated  $M_L/M_S$
ratio is substantially smaller than the experimental value
$R_{LS}=0.034$ \cite{Brewer}. Typically, $M_L$ is underestimated 
in LSDA due to the lack of orbital polarization. This leads to a smaller ratio $M_L/M_S$ and explains the disagreement  
with experiment.

\begin{table}[htbp]
  \caption{LSDA magnetic moments on Fe in Au (in units of $\mu_B$).}
  \label{tab:1}
  \begin{tabular}{cccc}
  \hline 
FeAu$_{15}$ & $M_S$ & $M_L$ & $M_L/M_S$ \\
\hline
FP-LAPW & 3.04  & 0.024     & 0.008 \\
VASP \cite{Costi}& 3.08 & 0.040 & 0.012 \\
TB-LMTO \cite{Frota}& 2.95 & 0.008& 0.003 \\
KKR-ASA \cite{Chadov}& -- & --    & 0.007 \\
\hline
  \end{tabular}
\end{table}

\begin{figure}[htbp]
\centerline{\hspace*{0.5cm}
\includegraphics[angle=270,width=10cm]{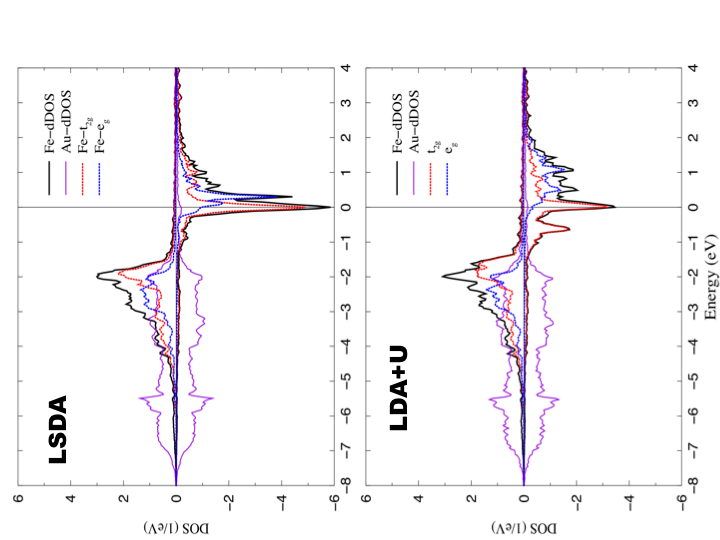}}
\caption{(Color online) The spin-resolved $d$-orbital DOS for Fe
  impurity in Au calculated with LSDA (top) and AMF-LDA+$U$, $U=3$ eV (bottom).
Also shown are $e_{g}$ and $t_{2g}$-like projected DOS for Fe.}
\label{fig:2}
\end{figure}

The calculated $d$-orbital density of states (DOS) for Fe atom and the
first nearest neighbor Au atoms are
shown in Fig.~\ref{fig:2}(top). LSDA yields fully occupied Fe spin-$\uparrow$
states that are hybridized with shallow Au $d$ states.  
The Fe spin-$\downarrow$ states are
substantially more localized. Analysis of the projected DOS
shows that the $e_g$-like ($d_{x^2-y^2} + d_{3z^2-r^2}$) states become practically fully spin-polarized while 
the $t_{2g}$-like ($d_{xy} + d_{zx} + d_{zy}$) states are only
partially polarized, see Fig.~\ref{fig:2}(top).

Now we turn to the LDA+$U$ calculations. 
We compare FLL and AMF variants of the rotationally-invariant LDA+$U$
method. The full local occupation
matrix with all spin off-diagonal components is preserved. The
double counting of the  non-spherical $d$-states contributions to the
LSDA and the LDA+$U$ parts of the potential is corrected. The
exchange $J = 0.9$ eV was used for Fe (Slater integrals $F_2=7.75$ eV, $F_4= 4.85$
eV). The Coulomb $U$ was varied from 3 eV to 5 eV. 

The spin $M_S$, orbital $M_L$ and dipole $M_D$ \cite{Wu}
$3d$ magnetic moments are given in Table~\ref{tab:2}  together with
the occupation of the Fe atom $d$ orbitals, $n_d$.  Both FLL and AMF flavors of
LDA+$U$ lead to an enhancement of $M_L$ with respect to the LSDA estimate.
It is due to  non-spherical Coulomb and exchange interactions which
are incorporated in LDA+$U$~\cite{bullmark} and cause an additional
orbital polarization to that induced by the spin-orbit coupling.
The  value of  $M_L$ increases with  increase of the Coulomb $U$. 
It is observed that the FLL double counting yields a substantially
stronger enhancement of $M_L$ than the AMF method.

There is also a substantial magnetic dipole moment $M_D$ formed on the
Fe impurity.  When the spin-orbit coupling is included and spin polarization is 
allowed, the initial cubic symmetry is broken and only the tetragonal symmetry remains.
This effect is rather small in LSDA. It becomes substantially enhanced in LDA+$U$ due to the additional
orbital polarization. This effect is visible on the AMF-LDA+$U$ DOS shown in Fig.~\ref{fig:2}(bottom). 
The main difference between LSDA and LDA+$U$ occurs in the
spin-$\downarrow$ channel for the
$t_{2g}$-like states; the $d_{xy}$ state
peels off from the $d_{zx}$ and $d_{zy}$ states and 
becomes occupied.

Experimental XMCD data are available for Fe impurity in Au~\cite{Brewer}. 
The measured value for $R_{LS}={M_L}/[M_S + 7M_D]=0.034$
is in a very good agreement with our AMF-LDA+$U$ results for
$U$ in the range between 3 eV and 4 eV.
On the basis of these
calculations we conclude that a reasonable value of  the
Coulomb $U$ for Fe impurity in Au host is $\approx$ 3 eV.

Our FLL results 
for $U = 5$ eV are fairly close to those of Ref.~\cite{Guo} where the
LDA+$U$ double counting was not specified.
In this case, the calculated $R_{LS}= 0.21$ exceeds the    
experimental XMCD value by almost an order of magnitude. Therefore, the
FLL-LDA+$U$ method does not satisfactorily describe the electronic
structure of Fe impurity in Au.

\begin{table}[htbp]
\caption{Magnetic moments (in $\mu_B$) and $3d$ occupation $n_d$ of
  the Fe impurity in Au host as a function of Coulomb $U$}
\label{tab:2}
\begin{center}
\begin{tabular}{cccc|cc}
\hline
 \multicolumn{1}{c}{FeAu$_{15}$}&\multicolumn{3}{c}{FLL} &\multicolumn{2}{c}{AMF} \\
\hline 
$U$ (eV)  &\multicolumn{1}{c}{3} &
  \multicolumn{1}{c}{4}&\multicolumn{1}{c|}{5}
  &\multicolumn{1}{c}{3} &
  \multicolumn{1}{c}{4}\\
 \hline
$M_S$            & 3.18   & 3.21 & 3.29& 2.94   & 2.90\\ 
$M_L$            & 1.24   & 1.36 & 1.44 &0.16   & 0.22 \\
$7M_D$         & 2.36   & 2.71 & 3.57 &2.16   & 2.35 \\
$R_{LS}$        & 0.23   & 0.23 & 0.21& 0.03   & 0.04 \\ 
$n_d$            & 6.00   & 6.00 & 5.97  & 6.03   & 6.03 \\
 \hline
\end{tabular}
\label{table2}
\end{center}
\end{table}

Both the LSDA and
LDA+$U$ methods yield broken-symmetry mean-field
solutions with non-zero $M_S$ and $M_L$. This is because the part of the Coulomb interaction 
treated in the Hartree--Fock-like approximation is transformed into the exchange splitting
field. This exchange field is of the order of a few eV (see Fig.~\ref{fig:2}) and by far exceeds
any imaginable external magnetic field. Thus, the  LDA+$U$ method, most probably, 
provides a reasonable description of the local-moment systems in (strong)
external magnetic fields.

When no external magnetic field is applied and no preferential direction for the
orientation of the moments exists, neither LSDA nor LDA+$U$ 
suffice. Recently, an attempt has been made to go beyond the static
mean-field approximation and to solve the SIAM for the Fe impurity in
Au employing the Hirsch--Fye quantum 
Monte Carlo method \cite{BoGu}. The authors used a simplified
three-orbital model 
with a diagonal Coulomb vertex and a spin-diagonal spin-orbit coupling
only. These simplifications make an estimate of the accuracy of the
quantitative results reported in Ref.~\cite{BoGu} difficult.

In order to deal with the electronic structure of the Fe impurity in
the absence of the external magnetic field, we apply
the finite-temperature ED method~\cite{JindraED} to the complete
five-orbital $d$ shell subject 
to the full spherically symmetric Coulomb interaction, spin-orbit coupling
and a cubic crystal field. The effective multi-orbital impurity Hamiltonian
can be written as \cite{Hewson}
\begin{align}
\label{eq:hamilt}
H - \mu N = & \sum_{km\sigma}
   \epsilon_k b^{\dagger}_{km\sigma} b_{km\sigma}
 + \sum_{m\sigma}\epsilon_d d^{\dagger}_{m \sigma}d_{m \sigma}
\nonumber \\
& + \sum_{mm'\sigma\sigma'} \bigl(\xi {\bf l}\cdot{\bf s} 
  + \Delta_{\rm CF}\bigr)_{m m'}^{\sigma \; \; \sigma'}
  d_{m \sigma}^{\dagger}d_{m' \sigma'}
\nonumber \\
& + \sum_{km\sigma} \Bigl( V_k
  d^{\dagger}_{m\sigma} b_{km\sigma} + \text{h.c.}
  \Bigr)
\\  
& + \frac{1}{2} \sum_{\substack {m m' m''\\  m''' \sigma \sigma'}} 
  U_{m m' m'' m'''} d^{\dagger}_{m\sigma} d^{\dagger}_{m' \sigma'}
  d_{m'''\sigma'} d_{m'' \sigma},
\nonumber
\end{align}
where $d^{\dagger}_{m \sigma}$ creates an electron in the $d$ shell and
$b^{\dagger}_{km\sigma}$ creates an electron in the ``bath''
which models those host-band states that hybridize with the
impurity $d$ shell. The bath is predominantly composed of $s$ and $p$ bands
of Au. The impurity-level position $\epsilon_d$ and the bath
energies $\epsilon_k$ are measured from the chemical potential $\mu$.
Parameters $\xi$ and $\Delta_{\rm CF}$ specify
the strength of the spin-orbit coupling and the size of the cubic
crystal field on the impurity. They are determined from LDA
calculations as $\xi=60$ meV and $\Delta_{\rm CF}=32$ meV.
The hybridization parameters $V_k$ and the bath energies
$\epsilon_k$ do not depend on $m$ and
$\sigma$, and thus the Hamiltonian
preserves orbital and spin angular momenta. This is a good
approximation for Fe in Au, since the lower-symmetry components of the
hybridization turn out to be considerably smaller than $\xi$ and
$\Delta_{\rm CF}$.

For the ED method to be applicable, the continuum of the bath states
is discretized. 
The parameters $\epsilon_d$, $\epsilon_k$ and $V_k$ are chosen so
that the impurity Green's function corresponding to the discretized
Eq.~\eqref{eq:hamilt} with $U=0$ approximates the
impurity Green's function from the LDA as closely as possible. Namely,
we require several lowest
moments of the respective densities of states to coincide,
$M_{n}^{\rm (SIAM)}=M_{n}^{\rm (LDA)}$, where 
$M_n=\int\epsilon^ng^d_0(\epsilon){\rm d}\epsilon
/\int g^d_0(\epsilon){\rm d}\epsilon$ \cite{georges1996}.
The integrals run over a  1 eV wide interval centered at the Fermi
level, which confines the LDA impurity resonance.
The actual values of the bath parameters
are $\epsilon_k^{\rm (I)}=80$ meV and $V_k^{\rm (I)}=220$ meV when the
index $k$ is restricted to a single value and the bath contains 10
spinorbitals (bath I: ``$d$+10 spinorbitals''). For a bath twice as large we get
$\epsilon_k^{\rm (II)}\in\{-310,\,340\}$ meV and 
$V_k^{\rm (II)}\in\{140,\,170\}$ meV (bath II: ``$d$+20 spinorbitals''). The
position of the impurity level $\epsilon_d$ obtained from this
procedure is subsequently shifted by a Hartree-like contribution in
order to maintain the LDA impurity occupation $n_d=6.18$ when the
local Coulomb term is introduced.

\begin{figure}[htbp]
\centerline{\includegraphics[angle=0,width=9.75cm]{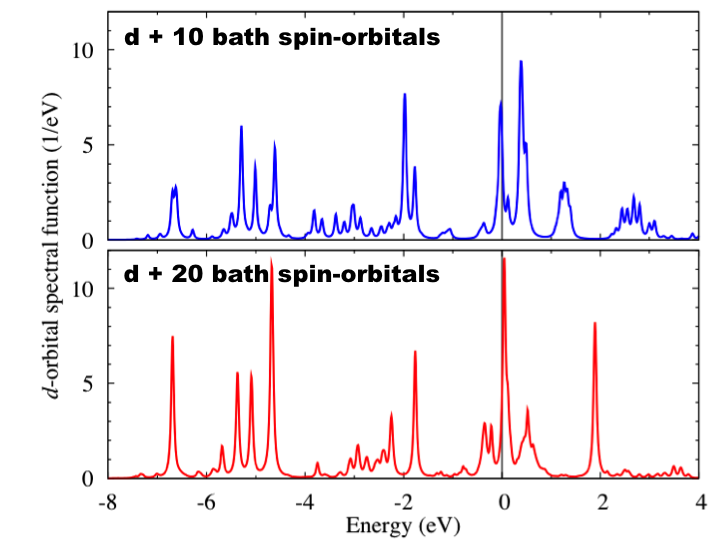}}
\caption{(Color online) $d$-electron spectral
  function of the impurity model of Eq.~\eqref{eq:hamilt} with $U= 3$
  eV and two variants of discrete bath: 10 bath spinorbitals (top)
  and 20 bath spinorbitals (bottom).} 
\label{fig:ED_dos}
\end{figure}

After the parameters of the discrete impurity model are set,
the band Lanczos 
method~\cite{ruhe1979,meyer1989} is utilized to determine the lowest
lying eigenstates of the many-body Hamiltonian and to calculate 
one-particle Green's function $G^d_{\rm SIAM}$. The resulting
$d$-orbital spectral function $\mathop{\rm Im}(G^d_{\rm SIAM})/\pi$
is shown in Fig.~\ref{fig:ED_dos} for the two models of the bath and for the
Coulomb interaction parameters $U=F_0=3$ eV and $J = 0.9$ eV ($F_2 =
7.75$ eV, $F_4= 4.85$ eV). The inverse temperature $\beta=500$
eV$^{-1}$ was used in these calculations. Although the details of the
spectral peaks depend somewhat on the particular choice of the bath,
the overall structure of the spectrum with peak(s) in the vicinity of
the Fermi level is preserved when the bath parameters are varied.
The spin $S=1.91$, orbital $L=2.21$ and total $J= 3.87$
moments are calculated for the $d$ shell from the expectation values
$\langle \hat X^2 \rangle=X(X+1)$, $X=S,L,J$.
Individual components of the moments, $\langle \hat S_z \rangle$ and
$\langle \hat L_z \rangle$, vanish so that the spin-orbital
symmetry is preserved and neither spin nor orbital polarization is
induced in the absence of the external magnetic field. 

Now we estimate the spin Hall angle from the skew scattering on the
impurity with a local magnetic moment.
Following Refs.~\cite{Engel,Guo}, we evaluate the spin Hall angle as
\begin{equation}
\gamma_S \cong \frac{12 \delta_1
  \bigl(\cos 2\delta^-_2 - \cos 2\delta^+_2\bigr)}{%
25 - 15\cos2\delta^+_2 - 10\cos2\delta^-_2}\,,
\label{eq:hall_angle}
\end{equation}
where $\delta_1$ is the $p$-wave phase shift which is assumed to be  
small for the non-resonant scattering, $|\delta_1| \cong 0.1$.
The $d$-wave phase shifts $\delta^+_2$ for $j=5/2$
and $\delta^-_2$ for $j=3/2$ are related to the
occupations $n_{3/2}$ and $n_{5/2}$
of the corresponding $3d$ subshells via the Friedel sum rule
$\delta_2^{(j)} = \pi n_j / (2j+1)$ \cite{Hewson}. The Hall angle
$\gamma_S$ vanishes when all $d$ orbitals are equally occupied
($\delta^-_2=\delta^+_2$), it grows as an increasing spin-orbit
coupling favors the occupation of the $j=3/2$ subshell,
and it eventually reaches a maximum $\gamma_S^{\rm (max)}=4\delta_1/5$
when the $j=3/2$ subshell is completely filled ($n_{3/2}=4$ and
$\delta^-_2=\pi$).

The occupation numbers $n_j$ and the Hall angle
$\gamma_S$ obtained for the Hamiltonian of Eq.~\eqref{eq:hamilt} are
listed in Table~\ref{tab:4} for the two bath 
models introduced earlier and for $U=3$ eV and 5~eV. Results of
the non-magnetic LDA calculation and of 
an atomic-like calculation without any bath orbitals are shown for
comparison. The angle $\gamma_S$ increases when the Coulomb $U$ is added,
and keeps growing with further increase of
$U$. For a fixed value of $U$, the Hall angle decreases
with increasing hybridization $V$ since the spin-orbit splitting in the
host band is negligible and the hybridization thus effectively reduces
the spin-orbit effects in the Fe $d$~shell.

Our results are consistent
with the  measurements of Fert {\em et al.}~\cite{Fert}  of the anomalous
Hall coefficient $\sim$ 0.01 in dilute 3$d$ noble metal alloys.
The angle $\gamma_S\approx 0.025$ we obtain is 50\% smaller
than the earlier theoretical estimate $\gamma_S=0.055$ by Gu {\em et
  al.} \cite{BoGu} and substantially
smaller than the ``giant'' $\gamma_S=0.11$ reported by Seki {\em et
  al.}~\cite{Seki}.
Note that the giant SHE interpretation of 
the experimental results in Ref.~\cite{Seki}
has been recently challenged also from
the experimental viewpoint~\cite{Mihajlovich}.    
\begin{table}[htdp]
\caption{Impurity occupations $n_j$ and the spin Hall
  angle $\gamma_S$ obtained with two different bath models for $U=3$ eV
  and 5 eV. Non-magnetic LDA calculation and an atomic-like
  calculation are shown for comparison.}
\begin{center}
\begin{tabular}{lccccc}
\hline
Model&$U$ (eV) &$n_d$&  $n_{3/2}$ & $n_{5/2}$ & $\gamma_S$ \\
\hline
LDA     &  --   & 6.18  &  2.62  & 3.55  & 0.008\\
bath I  &  3    & 6.18  &  2.68  & 3.50  & 0.011\\  
bath I  &  5    & 6.18  &  2.85  & 3.33  & 0.021\\  
bath II  &  3    & 6.18  &  2.94  & 3.24  & 0.026\\  
bath II  &  5    & 6.18  &  2.94  & 3.24  & 0.026\\  
no bath & 3--5  & 6.18  &  2.98  & 3.19  & 0.029\\
\hline
\end{tabular}
\end{center}
\label{tab:4}
\end{table}

%
To summarize, our calculations show that the AMF-LDA+$U$ method 
with the Coulomb $U$ around 3 eV reproduces very well 
the XMCD experimental data for Fe impurity in Au host. The calculated 
orbital moment at the Fe atom,  $M_L=0.16\mu_B$, is almost ten times smaller
than that reported by Guo {\em et al.}~\cite{Guo}. We explicitly show that
the reason for this difference is not only in the use of a smaller value of
$U$~\cite{BoGu}, but also in the appropriate choice of the LDA+$U$ flavor.
Furthermore, using the exact diagonalization of a multi-orbital impurity
model, we estimate the spin Hall angle due to the scattering on the Fe
impurity in the Au host as $\gamma_S \approx 0.025$. It is
substantially smaller than $\gamma_S = 0.11$
reported by Seki {\em et al.}~\cite{Seki}. 
We conclude that scattering off Fe impurities in Au does not yield a giant SHE.  

We acknowledge stimulating discussions with T. Jungwirth, J. Wunderlich, and 
J. Sinova, and financial support from Czech Republic
Grants GACR P204/10/0330, GAAV IAA100100912 and AV0Z10100520. 
J.K. acknowledges
support by the Alexander von Humboldt foundation.


\begin{thebibliography}{99}
\bibitem{Jungwirth2010}  
 Y. K. Kato, R. S. Myers, A. C. Gossard, and D. D. Awschalom, Science {\bf 306}, 1910 (2004);
 J. Wunderlich, B. Kaestner, J. Sinova, and T. Jungwirth, Phys. Rev. Lett. {\bf 94}, 047204 (2005). 
 
\bibitem{Kimura2007} 
S. O. Valenzuela and M. Tinkham, Nature {\bf 442}, 176 (2006). 


\bibitem{Seki} T. Seki, Y. Hasegawa, S. Mitani {\em et al.},
Nature Mater. {\bf 7}, 125 (2008).

\bibitem{Guo}  G. Guo, S. Maekawa, N. Nagaosa,  Phys. Rev. Lett. {\bf 102}, 036401 (2009).

\bibitem{Costi}T. Costi {\em et al.}, Phys. Rev. Lett. {\bf 102}, 056802 (2009).

\bibitem{Brewer} W. D. Brewer {\em et al.}, Phys. Rev. Lett. {\bf 93}, 077205 (2004).

\bibitem{BoGu} B. Gu {\em et al.}, Phys. Rev. Lett.  {\bf 105}, 086401 (2010).

\bibitem{Lixt95} A. I. Liechtenstein, V. I. Anisimov, and J. Zaanen,
  Phys. Rev. B {\bf 52}, R5467 (1995).

\bibitem{Hewson} A. C. Hewson, {\it The Kondo Problem to Heavy
Fermions}, Cambridge University Press, 1993.

\bibitem{shick01}  A. B. Shick and W. E. Pickett,
Phys. Rev. Lett.  {\bf 86}, 300 (2001).

\bibitem{vBH}  V. von Barth and L. Hedin, J. Phys. C {\bf 5}, 1629 (1972).

\bibitem{shick97} A. B. Shick, D. L. Novikov, and A. J. Freeman,
Phys. Rev. B {\bf 56}, R14259 (1997).

\bibitem{Frota} S. Frota-Pessoa, Phys. Rev. B {\bf 69}, 104401 (2004).

\bibitem{Chadov} S. Chadov {\em et al.}, Europhys. Lett. {\bf 82}, 37001 (2008).

\bibitem{Wu}  R. Wu and A. J. Freeman, Phys. Rev. Lett.  {\bf 73}, 1994
(1994).

\bibitem{bullmark} F. Bultmark, F. Cricchio, O. Granas, and L. Nordstrom,
Phys. Rev. B  {\bf 80}, 035121 (2009).  

\bibitem{JindraED} J. Koloren\v{c}, unpublished (2010).


\bibitem{georges1996} A. Georges {\em et al.}, Rev. Mod. Phys. {\bfseries 68}, 13 (1996).

\bibitem{ruhe1979} A. Ruhe, Math. Comput. {\bfseries 33}, 680 (1979).

\bibitem{meyer1989} H.-D. Meyer and S. Pal, J. Chem. Phys. {\bfseries 91}, 6195 (1989).


\bibitem{Engel} H.-A. Engel, B. I. Halperin, and E. I. Rashba,
Phys. Rev. Lett.  {\bf 95}, 166605 (2005).

\bibitem{Fert}  A. Fert, A. Friedrich, and A. Hamzic, J. Magn. Magn. Matter. {\bf 24}, 231 (1981).

\bibitem{Mihajlovich} G. Mihajlovic, J. E. Pearson, M. A. Garcia, {\em et al.}, Phys. Rev. Lett. {\bf 103}, 166601 (2009).
\end{thebibliography}
\end{document}